\RequirePackage[2020-02-02]{latexrelease}
\documentclass[aps,prb,twocolumn,floatfix,groupedaddress]
 {revtex4}
\usepackage[T1]{fontenc}
\usepackage[latin9]{inputenc}
\usepackage{graphicx}
\usepackage{amssymb}
\usepackage{amsmath}
\usepackage{color}



\begin{document}

\newcommand{\be}{\begin{equation}}
\newcommand{\ee}{  \end{equation}}
\newcommand{\ba}{\begin{eqnarray}}
\newcommand{\ea}{  \end{eqnarray}}
\newcommand{\ve}{\varepsilon}

\title{Effect of uniaxial magnetic anisotropy on charge transport in a junction with a precessing anisotropic molecular spin}

\author{Milena Filipovi\'{c}}
\affiliation{Institute of Physics Belgrade, University of Belgrade, Pregrevica 118, 11080 Belgrade, Serbia}

\date{\today}

\begin{abstract}
Anisotropic magnetic molecules can be employed to manipulate charge transport in molecular nanojunctions. Charge transport through a molecular orbital connected to two leads and exchange-coupled with a precessing anisotropic molecular spin in a constant magnetic field is studied here. Both the magnetic field and the uniaxial magnetic anisotropy parameter of the molecular spin modulate the total precession frequency. 
The precessing molecular magnetization drives inelastic tunneling processes between electronic quasienergy levels. The dc-bias voltages allow to unveil the quasienergy levels, Larmor frequency, and the anisotropy parameter, through characteristics of charge-transport measurements involving features such as steps, peaks and dips. Quantum interference effects between states connected with spin-flip events are reflected in the shot noise as peak-dip (dip-peak) features, resembling Fano-like resonance profiles, and are controlled by the anisotropy parameter and Larmor frequency. Under zero bias, the increase of the anisotropy parameter enables the decrease of the precession frequency or alters the precession direction, and shot noise is reduced. Furthermore, it is possible to adjust the anisotropy parameter to suppress the precession frequency, leading to the suppression of shot noise. The results show that in the given setup, the charge current and shot noise can be controlled by the magnetic anisotropy parameter of the molecular spin.
\end{abstract}

\maketitle

\section{Introduction}

Single-molecule magnets have gained much attention since the beginning of the new century due to the possibility to be used as constituent elements in spintronic devices for high-density information storage and quantum information processing.\cite{m1,m2,m3,m4,m5,m6} The key role in these applications plays the uniaxial magnetic anisotropy, characterized by parameter $D$, which leads to the bistability of the molecular spin states, with two degenerate ground states $\pm S$, separated by an energy barrier to spin reversal $DS^2$ (for integer spins) at low temperatures.\cite{m2,m3,m7,m8}
Depending on the sign of the anisotropy parameter, there are two types of uniaxial anisotropy: easy-axis ($D>0$), and easy-plane ($D<0$) anisotropy.\cite{m2} For successful applications in magnetic storage, the energy barrier needs to be enhanced,\cite{m9} but its increase cannot be accomplished by a simultaneous increase of the anisotropy parameter $D$ and the ground state spin $S$, and the only way to control the barrier height is to modulate the value of $D$.\cite{m10,m11,m12,m13} On the other hand, in-plane and small magnetic anisotropy is desirable for applications in quantum information processing.\cite{m14} In order to design magnetic molecules with desired characteristics, learning to control and manipulate the magnetic anisotropy parameter $D$ is essential.

Charge transport through magnetic molecules has been studied both theoretically\cite{m15,m16,bodecurrentinduced,t1,t2,we2013,t3,t31,t4,t5,t6,t7,t8} and experimentally.\cite{e1,e2,e3,manipulation1,e4,e5,e6,e7} The studies have addressed various phenomena, such as e.g., Kondo effect,\cite{kondo1,kondo2,kondo3,kondo4} Pauly spin blockade,\cite{spinblockade1,spinblockade2,spinblockade3} Coulomb blockade,\cite{e1,coulombblockade1,coulombblockade2,coulombblockade3} molecular magnetization 
dynamics in tunnel junctions\cite{bodecurrentinduced,Zhu4,Zhu5} or in contact with a superconducting lead,\cite{superconductor} and spin-dependent Seebeck and Peltier effect.\cite{seebeck1, seebeck2,seebeck3,seebeck4} The possibility to manipulate molecular magnetization by charge current has already been demonstrated experimentally.\cite{e2,e3,manipulation1,e4,e5,manipulation2} It has been theoretically predicted that exchange interactions between molecular magnets can be electrically controlled.\cite{t3,t31} Magnetic anisotropy can be varied and controlled by various means such as electrical current,\cite{t1,coulombblockade2,an1,an2,an3,an4} electric field,\cite{coulombblockade2,ele1,ele2,ele3,ele4} molecular mechanical stretching,\cite{stretching} and by ligand substitution.\cite{ligsub} High anisotropy barriers for spin reversal were observed in some isolated metal complexes, but due to their reactivity and instability, they are not suitable candidates to be exploited in magnetic storage.\cite{metal1,metal2,metal3,metal4} Also, new optical techniques of spin readout in single-molecule magnets have been investigated recently.\cite{opt1,opt2,opt3}

The nonequilibrium Green's functions technique\cite{Jauho1993,Jauho1994,JauhoBook} has been used to derive various characteristics of quantum transport through single molecules and molecular magnets, such as charge current, current-current correlations, spin current, inelastic transport, heat current, etc.\cite{t2,t3,t31,t4,seebeck4,innoiz1,neg1,neg2,t41,neg3} Charge-current noise in transport junctions arising from the discreteness of charge of conducting electrons is an exciting topic in nanophysics, since it can give us additional information about charge transport which is hidden from the current measurements.\cite{Blanter2024} Within the framework of nonequilibrium Green's functions technique, the effect of inelastic transport on shot noise has been studied,\cite{innoiz1,innoiz2,innoiz3,innoiz4,we2018} as well as current fluctuations in the transient regime.\cite{transient} It has been shown previously that a spin flip can lead to suppression\cite{innoiz2,noises} or enhancement\cite{noisee1,noisee2} of the shot noise. The shot noise has been employed to give information on e.g., energy of transmission channels,\cite{Blanter2024} fractional charges,\cite{fractional2024} and Cooper pairs.\cite{Cooper2024} Recently, the nonequilibrium noise due to temperature gradient at zero-bias voltage has become an active research topic.\cite{grad1,grad2,grad3,grad4,grad5} 

The goal of this paper is to theoretically study charge transport through a single electronic level that may be an orbital of a molecular magnet or belong to a nearby quantum dot, in the presence of a precessing anisotropic molecular spin in a constant external magnetic field, connected to electric contacts. The precession frequency is contributed by the Larmor precession frequency and a term with uniaxial magnetic anisotropy parameter of the molecular spin, and kept undamped by external means. The spin of the itinerant electron in the electronic level and the molecular spin are coupled via exchange interaction. The charge current and current noise are calculated by means of the Keldysh nonequilibrium Green's functions technique.\cite{Jauho1993,Jauho1994,JauhoBook} The shot noise of charge current is a result of the competition between correlations of currents with the same spins and correlations of currents with the opposite spins. Both elastic tunneling, driven by the dc-bias voltage, and inelastic tunneling involving a spin flip of the itinerant electrons, driven by the precessional motion of the anisotropic molecular spin, contribute to the charge transport. It is shown that by proper tuning of the uniaxial magnetic anisotropy parameter, the charge current and shot noise can be manipulated. The energies of the channels available for electron transport are Floquet quasienergies, which are obtained by the Floquet theorem.\cite{Floquet1,Floquet2,Floquet3,Floquet4} They are dependent on the magnetic anisotropy of the molecular spin and can be varied accordingly. Both charge current and shot noise are saturated for sufficiently large magnetic anisotropy parameter at nonzero bias-voltage conditions. Similarly to our previous work where the precessing molecular spin was isotropic,\cite{we2018} the peak-dip (dip-peak) features in the shot noise are manifestations of the quantum interference\cite{Fano2024,interf2024} between the states connected with inelastic tunneling, accompanied with a spin flip, here involving absorption (emission) of an energy which is linearly dependent on the uniaxial magnetic anisotropy parameter. 
Finally, the results show that the shot noise at zero-bias voltage conditions and zero temperature is reduced for chemical potentials of the metallic leads that enable inelastic tunnelling processes, if the magnetic anisotropy parameter increases, leading to the decrease of the total precession frequency or reversed direction of the molecular spin precession  with respect to the external magnetic field. Moreover, the magnetic anisotropy parameter can suppress the precession frequency of the molecular magnetization, leading to zero shot noise. 

The rest of the paper is arranged in the following way. The model setup of the system is described in Sec.~II. Theoretical formalism used to derive the results is given in Sec.~III. The expressions for the charge current and the current noise are calculated by means of the Keldysh nonequilibrium Green's functions technique.\cite{Jauho1993,Jauho1994,JauhoBook}
The results are shown and discussed in Sec.~IV, where the effects of the uniaxial magnetic anisotropy of the molecular spin on the charge-transport properties at zero temperture are analysed. This section is followed by Sec.~V in which the conclusions are presented. 

\section{Model Setup}

\begin{figure}
\includegraphics[width=8.5cm,keepaspectratio=true]{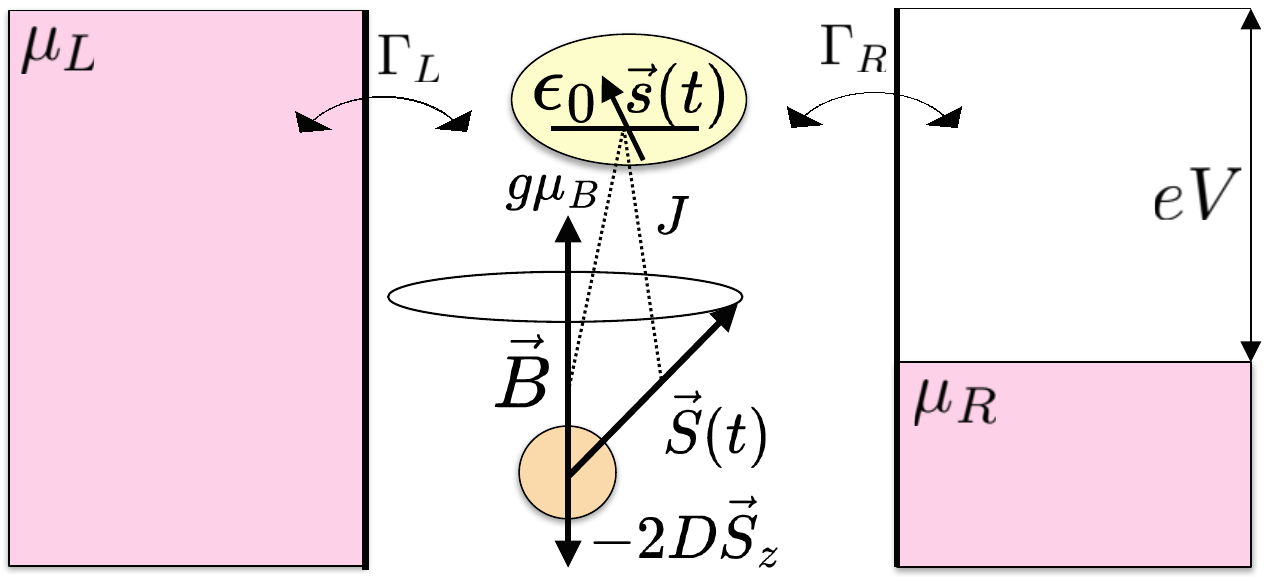}
\caption{Tunneling through a single molecular orbital with energy $\epsilon_{0}$ coupled to the anisotropic spin $\vec{S}(t)$ of a molecular magnet via  exchange interaction with the coupling constant $J$, in the presence of a magnetic field $\vec{B}$, connected to two metallic leads with chemical potentials $\mu_{L}$ and $\mu_R$. The applied bias voltage $eV=\mu_{L}-\mu_R$, with tunnel rates $\Gamma_L$ and $\Gamma_R$, and the uniaxial anisotropy constant of the molecular magnet $D$. The spin of the molecule precesses around the magnetic field axis with modified frequency $\omega=\omega_L-2DS_z$.}\label{fig: system}
\end{figure}

The junction under consideration consists of a single molecular orbital of a molecular magnet, in the presence of a precessing anisotropic molecular spin in a constant external magnetic field along $z$ axis, $\vec{B}=B\vec{e}_{z}$, coupled to two noninteracting metallic leads (see Fig.~1). The leads with chemical potentials $\mu_{L}$ (left) and $\mu_{R}$ (right) are unaffected by the magnetic field. The Hamiltonian describing the junction is given by $\hat{H}(t)=\hat{H}_{L}+\hat{H}_{R}+\hat{H}_{T}+\hat{H}_{\rm MO}(t)+\hat{H}_{S}$. Here, the Hamiltonian of lead $\xi=L,R$ can be written as $\hat{H}_\xi=\sum_{k,\sigma}\epsilon_{k\xi} \hat{c}^\dagger_{k\sigma\xi} \hat{c}_{k\sigma\xi}$. The subscript $\sigma=\uparrow,\downarrow=1,2=\pm 1$ denotes the spin-up or spin-down state of the electrons. The tunnel coupling between the molecular orbital and the leads is introduced by $\hat{H}_{T}=\sum_{k,\sigma,\xi}  [V_{k\xi}\hat{c}^\dagger_{k\sigma\xi} \hat{d}_{\sigma}+V^{\ast}_{k\xi}\hat{d}^\dagger_{\sigma} \hat{c}_{k\sigma\xi}]$, with matrix element $V_{k\xi}$. Here, $ \hat{c}^\dagger_{k\sigma\xi}(\hat{c}_{k\sigma\xi})$ and $ \hat{d}^\dagger_{\sigma} (\hat{d}_{\sigma})$ denote the creation (annihilation) operators of the electrons in the leads and the molecular orbital. The Hamiltonian of the molecular orbital is given by $\hat{H}_{\rm MO}(t)=\sum_{{\sigma}}\epsilon_{0}\hat{d}^\dagger_{\sigma} \hat{d}_{\sigma}+(g\mu_{B}/\hbar)\hat{\vec{s}}\vec {B}+J\hat{\vec{s}}\vec{S}(t)$, where the first term is the Hamiltonian of the noninteracting molecular orbital with energy $\epsilon_{0}$. The second term describes the electronic spin in the molecular orbital,  $\hat{\vec{s}}=(\hbar/2)\sum_{\sigma\sigma'}(\hat{\vec\sigma})_{\sigma\sigma'}\hat d^\dagger_\sigma\hat d_{\sigma'}$, in the presence of the magnetic field $\vec{B}$. The vector of the Pauli matrices is given by $\hat{\vec\sigma}=(\hat{\sigma}_x,\hat{\sigma}_y,\hat{\sigma}_z)^T$, while $g$ and $\mu_{B}$ are the $g$ factor of the electron and the Bohr magneton. The third term in the Hamiltonian of the orbital represents the exchange interaction between the electronic spin and spin of the molecule, where $J$ is the exchange coupling constant. The term $\hat{H}_{S}=(g\mu_{B}/\hbar){\hat{\vec S}}{\vec B}-D\hat{S^2_z}$ represents the Hamiltonian of the anisotropic molecular spin $\hat{\vec{S}}=\hat{S_x}\vec{e}_x+\hat{S_y}\vec{e}_y+\hat{S_z}\vec{e}_z$,  where $D$ is the uniaxial magnetic anisotropy parameter. It is presumed that $g$ factor of the molecular spin equals that of a free electron. 

Assuming that the spin of the molecular magnet is large and that it can be considered as a classical variable $\vec{S}$, with constant length $S=\textbar{\vec{S}}\textbar\gg\hbar$, where $\vec{S}=\langle\hat{\vec{S}}\rangle$ is the expectation value of the molecular spin operator, its dynamics is given by the Heisenberg equation of motion $\dot{\vec{S}} =\big \langle \dot{\hat{\vec{S}}} \big \rangle\ = ( i/\hbar)\big \langle \big [\hat{H},\hat{\vec{S}} \big ] \big \rangle$. Neglecting the quantum fluctuations and using external means, such as radiofrequency fields,\cite{Kittel} to compensate for the loss of the molecular magnetic energy due to its interaction with the itinerant electrons, so that the molecular spin dynamics remains unaffected by the exchange interaction, the equation
$\dot{\vec{S}}=(g\mu_{B}/\hbar)\vec{B}\times\vec{S}-2D\vec{S}_{z}\times\vec{S}$ is obtained.\cite{bodecurrentinduced} The molecular spin precesses around $z$ axis, with frequency $\omega=\omega_L-2DS_z$, where $\omega_{L}=g\mu_{B}B/\hbar$ is the Larmor precession frequency in the external magnetic field $\vec{B}$, while $-2DS_z$ is the contribution of the uniaxial anisotropy to the precession frequency $\omega$. The motion of the molecular spin is then given by $\vec S(t)=S_{\bot}\cos (\omega t)\vec e_x+S_{\bot}\sin (\omega t)\vec e_y+S_{z}\vec e_z$, where $\theta$ is the tilt angle between $z$ axis and $\vec{S}$, $S_{\bot}=S\sin(\theta)$ and $S_z=S\cos(\theta)$. Although the motion of the molecular spin is kept precessional externally,\cite{Kittel} the molecular magnet itself pumps charge current into the leads, thus affecting the transport properties of the junction.
In the rotating reference frame attached to the end point of the molecular spin, the Hamiltonian of the molecular orbital $\hat{\tilde{H}}_{\rm MO}$ is time-independent and can be written as $\hat{\tilde{H}}_{\rm MO}=\hat{U}\hat{H}_{\rm MO}\hat{U}^\dagger+i\hbar(\partial_{t}\hat{U})\hat{U}^\dagger$, where $\hat{U}$ is the unitary operator $\hat{U}(t)=e^{i\omega t\hat{s}_{z}/\hbar}$. The orbital Hamiltonian $\hat{\tilde{H}}_{\rm MO}$ can be further expressed as
$\hat{\tilde{H}}_{\rm MO}=\sum_{{\sigma}}\epsilon_{0}\hat{d}^\dagger_{\sigma} \hat{d}_{\sigma}+ (2D+J)S_{z}\hat{s}_{z}+JS_{\bot}\hat{s}_x$.

Separation between a molecular level and a localized molecular spin can be obtained in e.g., molecular magnets containing atoms of transition metals or rare earth elements,\cite{t4,an2,Godfrin} since the molecular spin here is composed of localized d or f orbitals belonging to transition metal or rare earth element. The s and p orbitals in the molecular ligands form the localized molecular level, i.e., the highest occupied or lowest unoccupied molecular orbital. If we take into account that the Coulomb interaction in the s and p orbitals in the ligands can be considered negligible, then the on-site Coulomb repulsion can be neglected.

\section{Theoretical Formalism}
\subsection{Charge Current}
The charge-current operator of the contact $\xi$ is given by
the Heisenberg equation \cite{JauhoBook,Jauho1994} 
\begin{equation}
	\hat{I}_{\xi}(t)=-e\frac{d\hat{N}_{\xi}}{dt}=-e\frac{i}{\hbar}[\hat{H},\hat{N}_{\xi}],\label{eq: commutator}
\end{equation}
where $\hat{N}_{\xi}=\sum_{{k,\sigma}}
\hat{c}^\dagger_{k\sigma\xi}\hat{c}_{k\sigma\xi}$ represents the charge occupation number operator of the contact $\xi$, while $[\ , \ ]$ denotes the commutator.
The average charge current from the lead $\xi$ to the molecular orbital is then given by 
\begin{equation}
	I_{\xi}(t) =-e\bigg \langle \frac{d}{dt} \hat{N}_{\xi} \bigg \rangle\ =  -e\frac{i}{\hbar} 
	\big \langle \big [\hat{H},\hat{N}_{\xi} \big ] \big \rangle.
\end{equation}
Using the Keldysh nonequilibrium Green's functions technique, the charge current can be calculated as \cite{Jauho1994, JauhoBook}
\begin{align}
	\label{eq: general current}
	I_{\xi}{(t)} =&{2e}\ {\rm Re}\int dt^\prime {\rm Tr}\big\{ {\hat{G}}^{r} {(t,t^\prime)}{\hat{\Sigma}}^{<}_{\xi}(t^\prime,t)\nonumber\\
	& \quad \quad \quad\quad \quad\quad +{\hat{G}}^{<} {(t,t^\prime)}{\hat{\Sigma}}^{a}_{\xi}(t^\prime,t) \big\},
\end{align}
in units in which $\hbar=e=1$. 
The retarded, advanced, lesser and greater self-energies from the tunnel coupling between the molecular orbital
and contact $\xi$ are denoted by $\hat{\Sigma}^{r,a,<,>}_{\xi}(t,t^\prime)$. Their matrix elements are diagonal in the electron spin space with respect to the basis of the eigenstates of $\hat{s}_{z}$, with nonzero matrix elements given by 
$\Sigma^{r,a,<}_{\xi}(t,t')=\sum_{{k}}
V_{k\xi}^{\phantom{\ast}}g^{r,a,<,>}_{k\xi}{(t,t')}V^{\ast}_{k\xi}$,
where $g^{r,a,<,>}_{k \xi}{(t,t')}$ represent the retarded, advanced, lesser and greater Green's functions of the electrons in the lead $\xi$. The Green's functions of the electrons in the molecular orbital are given by
$\hat{G}^{r,a,<,>}(t,t^\prime)$,
with matrix elements 
$G^{r,a}_{\sigma\sigma^\prime} {(t,t^\prime)}=\mp i\theta(\pm t \mp t^\prime)\langle\{\hat{d}_{\sigma}{(t)},
\hat{d}^\dagger_{\sigma^\prime} {(t^\prime)}\}\rangle$, while $G^<_{\sigma\sigma^\prime} (t,t^\prime)= i \langle \hat{d}^\dagger_{\sigma^\prime} (t^\prime) \hat{d}_\sigma(t)\rangle$ and
$G^>_{\sigma\sigma^\prime} (t,t^\prime)= -i \langle \hat{d}_{\sigma} (t) \hat{d}^\dagger_{\sigma^\prime}(t^\prime)\rangle$, with the anticommutator denoted as $\{\cdot ,\cdot\}$.
Applying the double Fourier transformations in Eq.~(3), one obtains 
\begin{align}\label{eq: struja}
	I_{\xi}{(t)}=&\,-2e\Gamma_{\xi}{\rm Im}\int\frac{d\epsilon}{2\pi}\int\frac{d\epsilon^\prime}{2\pi} e^{-i(\epsilon-\epsilon^\prime)t}\nonumber\\
	&\times{\rm Tr}\bigg\{f_{\xi}(\epsilon^{\prime}){\hat{G}}^{r} {(\epsilon,\epsilon^{\prime})+ {\frac{1}{2}{\hat{G}}^{<}} {(\epsilon,\epsilon^{\prime})}}\bigg\},
\end{align}
where the tunnel coupling between the molecular orbital and contact $\xi$, $\Gamma_{\xi}(\epsilon)=2\pi\sum_{{k}}\lvert V_{k\xi}\rvert^{2}\delta (\epsilon-\epsilon_{k\xi})$, is energy independent in the wide-band limit and considered constant. The Fermi-Dirac distribution of the electrons in the lead $\xi$ is given by $f_{\xi}(\epsilon)=[e^{(\epsilon-\mu_{\xi})/k_{B}T}+1]^{-1}$,
with $k_{B}$ the Boltzmann constant and $T$ the temperature.

The matrix elements of the retarded Green's function $\hat{G}^{r}(t,t^{\prime})$ of the electrons in the molecular orbital, can be obtained by applying Dyson's expansion and analytic continuation rules,\cite{JauhoBook}
\begin{equation}\label{eq: Dyson_Green}
	\hat{G}^r(t,t^\prime)=\hat{G}^{0r}(t-t^\prime)+\int dt_{1}\hat{G}^{0r}(t-t_1)\hat{H}^{\prime}(t_{1})\hat{G}^r(t_{1},t^\prime).
\end{equation}
Here, $\hat{G}^{0r}(t-t^\prime)$ is the retarded Green's function of the electrons in the orbital in the presence of only the static component of the molecular spin $S_{z}$ along the axis of the external magnetic field. It can be found using the equation of motion technique,\cite{Bruus} and after applying the Fourier transformations, one obtains
$\hat{G}^{0r}(\epsilon)=[\epsilon-\epsilon_{0}-\Sigma^{r}-\hat{\sigma}_{z} (g\mu_{B}B+J S_{z})/2]^{-1}$,\cite{Guo,bodecurrentinduced} where $\Sigma^{r,a}=\mp i\Gamma/2$ and $\Gamma=\sum_{\xi}\Gamma_{\xi}$. In the second term of Eq.~(5), $\hat{H}^{\prime}(t)=\gamma(e^{-i\omega t}\hat{d}^{\dagger}_{\uparrow}\hat{d}_{\downarrow}+e^{i\omega t}\hat{d}^{\dagger}_{\downarrow}\hat{d}_{\uparrow})$ is the off diagonal part of the Hamiltonian, representing the exchange interaction between the electronic spin in the orbital and the rotational component of the molecular spin, with $\gamma=JS_{\bot}/2$. The double Fourier transforms of the matrix elements of $\hat{G}^r(t,t^{\prime})$ read\cite{Guo,we2016}
\begin{align}\label{eq: greenn}
	G^{r}_{\sigma\sigma}(\epsilon,\epsilon^\prime)&
	=\frac{2\pi \delta(\epsilon-\epsilon^\prime)G^{0r}_{\sigma\sigma}(\epsilon)}{1-\gamma^{2}G^{0r}_{\sigma\sigma}(\epsilon)
		{G^{0r}_{-\sigma-\sigma}(\epsilon_{\sigma})}},\\
	\label{eq: greenn1}	G^{r}_{\sigma-\sigma}(\epsilon,\epsilon^\prime)&
	=\frac{2\pi\gamma\delta(\epsilon_{\sigma}-\epsilon^{\prime})G^{0r}_{\sigma\sigma}(\epsilon)
		G^{0r}_{-\sigma-\sigma}(\epsilon_{\sigma})}{
		1-\gamma^{2}G^{0r}_{\sigma\sigma}(\epsilon)
		G^{0r}_{-\sigma-\sigma}(\epsilon_{\sigma})},
\end{align}
where the abbreviation $\epsilon_{\sigma}=\epsilon-\sigma\omega=\epsilon-\sigma (\omega_L-2DS_{z})$ is used.
Applying the double Fourier transformations to the Keldysh equation\cite{JauhoBook}
\begin{equation}\label{eq: greenn2}
	\hat{G}^{<,>}(t,t^\prime)=\int dt_{1}dt_{2}\hat{G}^{r}(t,t_{1})\hat{\Sigma}^{<,>}(t_{1}-t_{2})\hat{G}^{a}(t_{2},t^{\prime}),
\end{equation}
the lesser and greater Green's functions can be calculated as $\hat{G}^{<,>}(\epsilon,\epsilon^\prime)=\int d\epsilon^{\prime\prime}\hat{G}^{r}(\epsilon,\epsilon^{\prime\prime})\hat{\Sigma}^{<,>}(\epsilon^{\prime\prime})\hat{G}^{a}(\epsilon^{\prime\prime},\epsilon^\prime)/2\pi$, where ${\Sigma}^{<}(\epsilon)=i\sum_{\xi}\Gamma_{\xi}f_{\xi}(\epsilon)$, ${\Sigma}^{>}(\epsilon)=i\sum_{\xi}\Gamma_{\xi}(f_{\xi}(\epsilon)-1)$ and $\hat{G}^{a}(\epsilon,\epsilon^\prime)=[\hat{G}^{r}(\epsilon^\prime,\epsilon)]^\dagger$.

Finally, using Eqs.~(4)--(8), the average charge current from the contact $\xi$ can be written as
\begin{align}\label{eq: strujae}
	I_{\xi}&=\frac{e\Gamma_{\xi}\Gamma_{\zeta}}{\hbar}\int\frac{d\epsilon}{2\pi}[f_{\xi}(\epsilon)-f_{\zeta}(\epsilon)]\nonumber\\
	&\times\sum_{\substack{\sigma\sigma' \\ \sigma\neq\sigma'}}
	\frac{\lvert G^{0r}_{\sigma\sigma}(\epsilon)\rvert^{2}[1+\gamma^{2}\lvert G^{0r}_{\sigma'\sigma'}(\epsilon+\sigma'\omega_{L}-2\sigma' DS_{z})\rvert^2]}{\lvert 1-\gamma^{2} G^{0r}_{\sigma\sigma}(\epsilon)G^{0r}_{\sigma'\sigma'}(\epsilon+\sigma'\omega_{L}-2\sigma' DS_{z})\rvert^2},
\end{align}
with $\xi\neq\zeta$. Using the spin-resolved charge currents $I^{\sigma}_{\xi}$, the charge current can be written as $I_{\xi}=I^{\uparrow}_{\xi}+I^{\downarrow}_{\xi}$. In the limit $DS_{z}\ll\omega_{L}$, Eq.~(9) reduces to the previously calculated expression for the charge current.\cite{we2018}

\subsection{Density of States in the Molecular Orbital}

The positions of the resonant transmission channels available for electron transport in the molecular orbital can be obtained from the density of states in the orbital 
\begin{equation}
	\rho(\epsilon)=-\frac{1}{\pi}\sum_{\sigma=\pm 1}{\rm Im}\bigg\{\frac{G^{0r}_{\sigma\sigma}(\epsilon)}{1-\gamma^{2}G^{0r}_{\sigma\sigma}(\epsilon)G^{0r}_{-\sigma-\sigma}(\epsilon_{\sigma})}\bigg\}.
\end{equation} 
Taking into account that the Hamiltonian of the molecular orbital is a periodic function of time $\hat{H}_{\rm MO}(t)=\hat{H}_{\rm MO}(t+\mathcal{T})$, with $\mathcal{T}=2\pi/\omega$, the energies of the molecular levels at which the resonant transmission channels are located, can be calculated as Floquet quasienergies $\epsilon_i$, $i\in\{1,2,3,4\}$, using the Floquet theorem.\cite{Floquet1,Floquet2,Floquet3,Floquet4} They are given by
\begin{align}
	\epsilon_{1,3}=\epsilon_{0}-\frac{\omega_{L}}{2}+&DS_{z}\pm\sqrt{D(D+J)S^{2}_{z}+\bigg (\frac{JS}{2}\bigg )^2},\label{eq: epsilon_1,3}\\
	\epsilon_{2,4}=\epsilon_{0}+\frac{\omega_{L}}{2}-&DS_{z}\pm\sqrt{D(D+J)S^{2}_{z}+\bigg (\frac{JS}{2}\bigg )^2}.\label{eq: epsilon_2,4}
\end{align}
The derivation of the Floquet quasienergies is presented in the Appendix. Note that $\epsilon_{2}=\epsilon_{1}+\omega$ and $\epsilon_{4}=\epsilon_{3}+\omega$. Some of the spin-flip absorption and emission processes are shown in Fig.~2, for $\omega>0$.
In the molecular orbital, an electron with energy $\epsilon_{1}(\epsilon_{2})$ or $\epsilon_{3}(\epsilon_{4})$ can absorb (emit) an energy equal to one energy quantum $\omega$ and flip its spin, due to the exchange interaction with the precessing component of the anisotropic molecular spin, ending up in the quasienergy level with energy $\epsilon_{2}(\epsilon_{1})$ or $\epsilon_{4}(\epsilon_{3})$, and then tunnelling into either lead, so that the $\downarrow$ state with quasienergy $\epsilon_{1}(\epsilon_{3})$ is coupled with the $\uparrow$ state with quasienergy $\epsilon_{2}(\epsilon_{4})$. An inelastic spin-flip process is presented in Fig.~2(a). Here, an electron tunnels from the left lead into the $\downarrow$ level $\epsilon_{1}$ (or $\epsilon_{3}$) of the molecular orbital, flips its spin and absorbs an energy quantum $\omega$. Then it tunnels into either lead.

\begin{figure}
	\includegraphics[width=8.7cm,keepaspectratio=true]{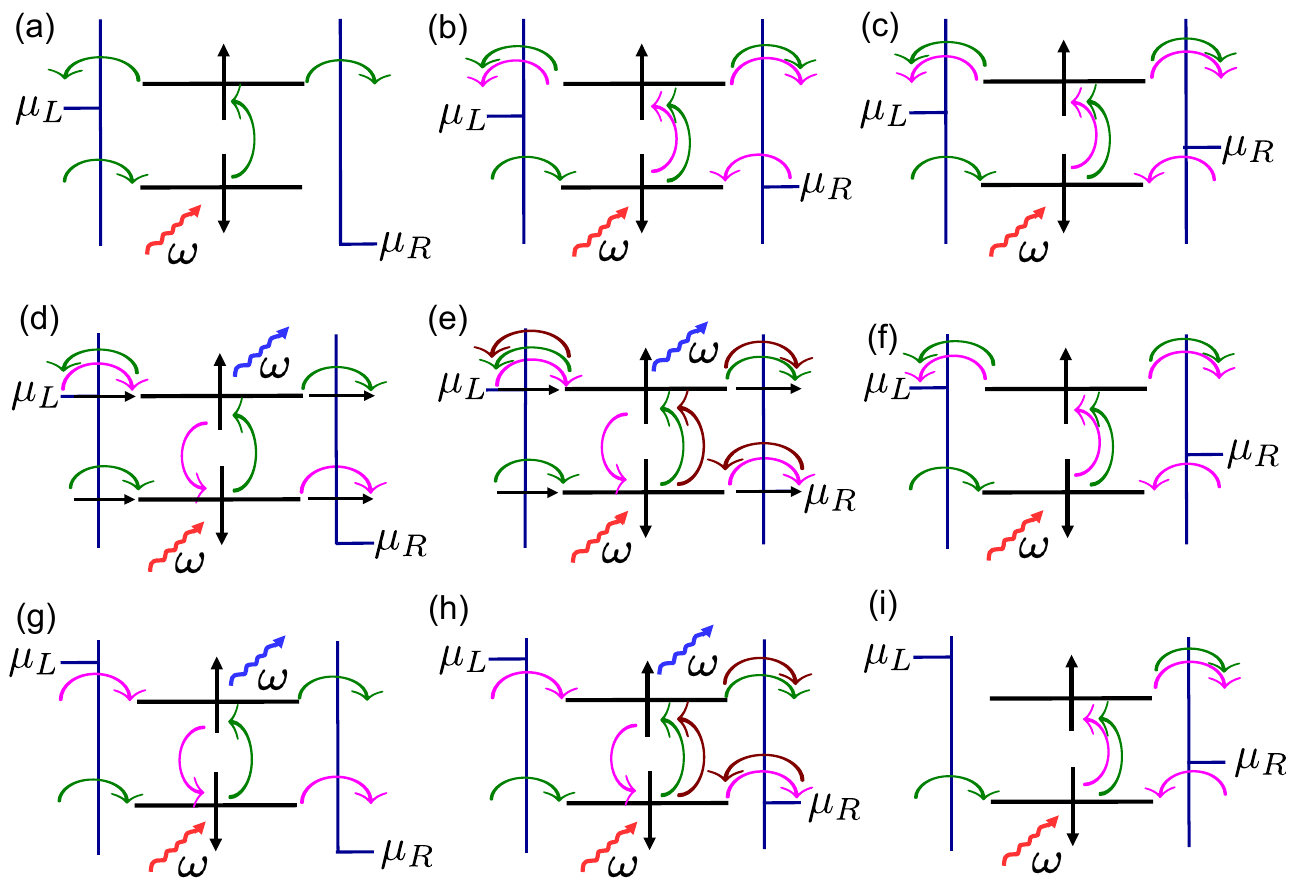}
	\caption{Inelastic spin-flip processes between quasienergy levels of the molecular orbital in the presence of the precessing anisotropic molecular spin with frequency $\omega=\omega_{L}-2DS_{z}$, for different positions of the levels with respect to chemical potentials $\mu_{L}$ and $\mu_{R}$. Different processes are represented with different colors. In (d) and (e), the elastic tunneling processes are represented with horizontal black arrows. For $\omega>0$, a spin $\downarrow$ ($\uparrow$) electron absorbs (emits) an energy quantum $\omega$ and flips its spin due to the exchange interaction with the precessing component of the molecular spin.}\label{fig: tunnelling}
\end{figure}
 
\subsection{Noise of Charge Current}

Additional properties of the charge transport in the junction can be obtained by analysing the charge-current noise. In view of the fact that the nonzero commutator in Eq.~(1) is generated by the tunneling Hamiltonian $\hat{H}_T$, the charge current operator $\hat{I}_{\xi}(t)$ can be written as 
\begin{equation}
	\hat{I}_{\xi}(t)=e\frac{i}{\hbar}\sum_{\sigma}\hat{I}_{\xi\sigma}(t),\label{eq: sigmacomponents}
\end{equation}
with the operator component $\hat{I}_{\xi\sigma}(t)$ given by
\begin{equation}
	\hat{I}_{\xi\sigma}(t)=\sum_{k}[V_{k\xi}\hat{c}^{\dag}_{k\sigma\xi}(t)\hat{d}_{\sigma}(t)-V^{*}_{k\xi}\hat{d}^{\dag}_{\sigma}(t)\hat{c}_{k\sigma\xi}(t)].
\end{equation}
The fluctuation operator of the charge current in contact $\xi$ is given by
\begin{equation}
	\delta\hat{I}_{\xi}(t)=\hat{I}_{\xi}(t)-\langle\hat{I}_{\xi}(t)\rangle.\label{eq: correlation}
\end{equation}
The correlation between fluctuations of currents in leads ${\xi}$ and ${\zeta}$, known as nonsymmetrized charge-current noise is written as\cite{JauhoBook,Blanter2024}
\begin{equation}
	S_{\xi\zeta}(t,t')=\langle\delta\hat{I}_{\xi}(t)\delta\hat{I}_{\zeta}(t')\rangle,\label{eq: 5.4}
\end{equation}	
while the symmetrized noise is defined as\cite{JauhoBook,Blanter2024} 
\begin{equation}
	S_{\xi\zeta S}(t,t')=\frac{1}{2}\langle\{\delta\hat{I}_{\xi}(t),\delta\hat{I}_{\zeta}(t')\}\rangle.\label{eq: 5.4s}
\end{equation}	
With the help of Eqs.~(13)--(15), one obtains the nonsymmetrized noise as 
\begin{equation}
	S_{\xi\zeta}(t,t')=\sum_{\sigma\sigma'}
	S^{\sigma\sigma'}_{\xi\zeta}(t,t'),\label{eq: 5.6}
\end{equation}
with $S^{\sigma\sigma'}_{\xi\zeta}(t,t')=(-e^{2}/\hbar^{2})\langle\delta\hat{I}_{\xi\sigma}(t)\delta\hat{I}_{\zeta\sigma'}(t')\rangle$ representing the correlation between fluctuations of spin-resolved charge currents $I_{\xi\sigma}$ and $I_{\zeta\sigma'}$.
Applying Wick's theorem\cite{Wick2024} and Langreth analytical continuation rules,\cite{Langreth2024} the correlation functions $S^{\sigma\sigma'}_{\xi\zeta}(t,t')$ introduced in Eq.~(18) can be calculated.\cite{JauhoBook,transient} Using the Fourier transforms of Green's functions $G^{r,a,<,>}_{\sigma\sigma'}(\epsilon,\epsilon')$ and self-energies $\Sigma^{r,a,<,>}_{\xi}(\epsilon)$, 
the formal expression for the nonsymmetrized noise of charge current in standard coordinates $t$ and $t^{\prime}$, obtained previously,\cite{JauhoBook,we2018,transient} becomes
\begin{widetext}
\begin{align}
	S_{\xi\zeta}(t,t')=-\frac{e^2}{\hbar^2}\sum_{\sigma\sigma'}\bigg\{\int&\frac{d\epsilon_{1}}{2\pi}\int \frac{d\epsilon_{2}}{2\pi}\int \frac{d\epsilon_{3}}{2\pi}	\int \frac{d\epsilon_{4}}{2\pi}e^{-i(\epsilon_{1}-\epsilon_{2})t}e^{i(\epsilon_{3}-\epsilon_{4})t'}\nonumber\\
	\times&\big\{[G^{r}_{\sigma\sigma'}(\epsilon_{1},\epsilon_{3})\Sigma^{>}_{\zeta}(\epsilon_{3})+2G^{>}_{\sigma\sigma'}(\epsilon_{1},\epsilon_{3})\Sigma^{a}_{\zeta}]
	[G^{r}_{\sigma'\sigma}(\epsilon_{4},\epsilon_{2})\Sigma^{<}_{\xi}(\epsilon_{2})+2G^{<}_{\sigma'\sigma}(\epsilon_{4},\epsilon_{2})\Sigma^{a}_{\xi}]\nonumber\\
	&+[\Sigma^{>}_{\xi}(\epsilon_{1})G^{a}_{\sigma\sigma'}(\epsilon_{1},\epsilon_{3})+2G^{>}_{\sigma\sigma'}(\epsilon_{1},\epsilon_{3})\Sigma^{r}_{\xi}]
	[\Sigma^{<}_{\zeta}(\epsilon_{4})G^{a}_{\sigma'\sigma}(\epsilon_{4},\epsilon_{2})+2G^{<}_{\sigma'\sigma}(\epsilon_{4},\epsilon_{2})\Sigma^{r}_{\zeta}]\nonumber\\
	&+4\Sigma^{r}_{\xi}\Sigma^{a}_{\zeta}G^{>}_{\sigma\sigma'}(\epsilon_{1},\epsilon_{3})G^{<}_{\sigma'\sigma}(\epsilon_{4},\epsilon_{2})\big\}\nonumber\\
	&-\delta_{\xi\zeta}\delta_{\sigma\sigma'}\int \frac{d\epsilon_{1}}{2\pi}\int \frac{d\epsilon_{2}}{2\pi}\int \frac{d\epsilon_{3}}{2\pi}\nonumber\\
	&\hspace{0.4cm}\times\big\{e^{-i(\epsilon_{1}-\epsilon_{3})t}e^{i(\epsilon_{2}-\epsilon_{3})t'}G^{>}_{\sigma\sigma'}(\epsilon_{1},\epsilon_{2})\Sigma^{<}_{\xi}(\epsilon_{3})\nonumber\\
	&\hspace{0.4cm}+e^{-i(\epsilon_{1}-\epsilon_{3})t}e^{i(\epsilon_{1}-\epsilon_{2})t'}\Sigma^{>}_{\xi}(\epsilon_{1})G^{<}_{\sigma'\sigma}(\epsilon_{2},\epsilon_{3})\big\}\bigg\}.\label{eq: neznam26maj}
\end{align}	
\end{widetext}
\begin{figure*}
	\includegraphics[height=5.65cm,keepaspectratio=true]{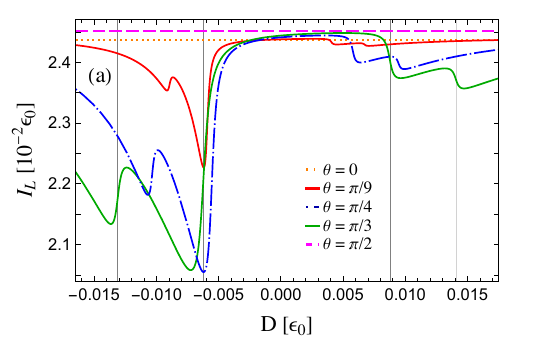}  
	\includegraphics[height=5.65cm,keepaspectratio=true]{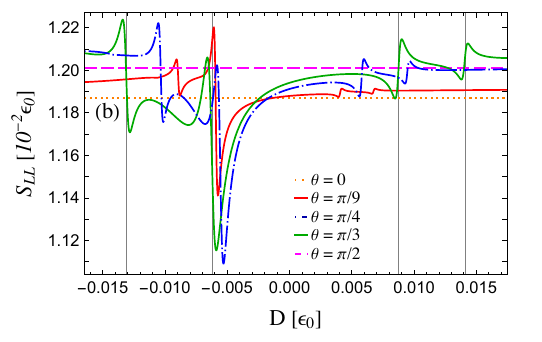}    
	\caption{(a) Charge current $I_L$ and (b) autocorrelation shot noise $S_{LL}$ as functions of the uniaxial magnetic anisotropy parameter $D$ for different tilt angles $\theta$, at zero temperature, with $\vec{B}=B\vec{e}_z$. The chemical potentials of the leads are equal to $\mu_{L}=\nobreak2.5\,\epsilon_0$ and $\mu_{R}=0$. The other parameters are set to $\Gamma=\nobreak0.05\, \epsilon_{0},\,\Gamma_{L}=\Gamma_{R}=\Gamma/2,\,\omega_L=\nobreak0.5\,\epsilon_{0},\, J=\nobreak0.01\,\epsilon_{0}$, and $S=100$. Grid lines for $\theta=\pi/3$ (green line) are positioned at $D=-0.01312\,\epsilon_0$ ($\mu_{L}=\epsilon_2$), $D=-0.00625\,\epsilon_0$ ($\mu_{R}=\epsilon_3$), $D=0.00875\,\epsilon_0$ ($\mu_{R}=\epsilon_4$), and $D=0.01406\,\epsilon_0$ ($\mu_{L}=\epsilon_1$).}\label{fig: current_D}
\end{figure*}
The resulting nonsymmetrized noise depends only on the time difference $\tau =t-t'$, and its power spectrum is given by
\begin{equation}
	S_{\xi\zeta}(\Omega)=\int d\tau e^{i\Omega\tau}S_{\xi\zeta}(\tau),\label{eq: 5.17}
\end{equation}
while the symmetrized noise spectrum reads
\begin{equation}
	S_{\xi\zeta S}(\Omega)=\frac{1}{2}[S_{\xi\zeta}(\Omega)+S_{\zeta\xi}(-\Omega)].\label{eq: 5.177}
\end{equation}
The individual noise components, $S^{\sigma\sigma'}_{\xi\zeta}(\Omega)$ and symmetrized $S^{\sigma\sigma'}_{\xi\zeta S}(\Omega)=\frac{1}{2}[S^{\sigma\sigma'}_{\xi\zeta}(\Omega)+S^{\sigma'\sigma}_{\zeta\xi}(-\Omega)]$ can be calculated using Eqs.~(18)--(21), and represent correlations between fluctuations of charge currents with the same spins for $\sigma=\sigma'$, or different spins for $\sigma\neq\sigma'$.
The charge current given by Eq.~(9) is conserved, implying that the zero-frequency ($\Omega=0$) noise power satisfies the relations $S_{LL}(0)=S_{RR}(0)=-S_{LR}(0)=-S_{RL}(0)$. In experimental configurations zero-frequency noise power is standardly measured. In the remainder of this article, the noise power $S_{LL}=S_{LL}(0)=S_{LLS}(0)$ at zero temperature will be discussed, as in this particular case it is contributed only by the shot noise, while thermal noise vanishes. The elastic tunnelling events are induced by the bias-voltage. The inelastic tunnelling events, each involving a spin flip and absorption (emission) of an energy quantum $\omega$, are induced by the exchange interaction between the spins of the conduction electrons and the precessing component of the anisotropic molecular spin.
The second term in Eq.~(19) involving $\delta_{\sigma\sigma'}$ vanishes for $\sigma\neq\sigma'$. Hence, the shot noise $S_{LL}$ is the result of the competition between  positive correlations of currents with the same spins $S^{\uparrow\uparrow}_{LL}=S^{\uparrow\uparrow}_{LLS}$ and $S^{\downarrow\downarrow}_{LL}=S^{\downarrow\downarrow}_{LLS}$, and negative correlations of currents with opposite spins, induced by spin-flip events, $S^{\uparrow\downarrow}_{LL}$ and $S^{\downarrow\uparrow}_{LL}$, which are a complex conjugate pair $S^{\downarrow\uparrow}_{LL}=(S^{\uparrow\downarrow}_{LL})^{*}$, with $S^{\uparrow\downarrow}_{LLS}=S^{\downarrow\uparrow}_{LLS}$.

\section{Results}

\begin{figure} 
	\includegraphics[height=5.65cm,keepaspectratio=true]{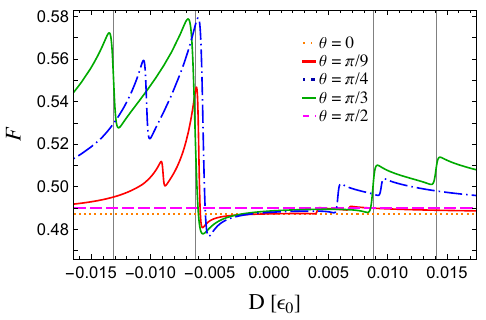}   
	\caption{Fano factor $F$  as a function of the uniaxial anisotropy parameter $D$. The plots are obtained for different tilt angles $\theta$ at zero temperature, with $\vec{B}=B\vec{e}_z$. The chemical potentials of the leads are equal to $\mu_{L}=\nobreak2.5\,\epsilon_0$ and $\mu_{R}=0$. The other parameters are set to $\Gamma=\nobreak0.05\, \epsilon_{0},\,\Gamma_{L}=\Gamma_{R}=\Gamma/2,\,\omega_L=\nobreak0.5\,\epsilon_{0},\, J=\nobreak0.01\,\epsilon_{0}$, and $S=100$.}\label{fig: Fano_D}
\end{figure}

Now we analyse the behaviour of the charge current $I_L$, zero-frequency noise power $S_{LL}$, and Fano factor $F=S_{LL}/e|I_L|$ as functions of the uniaxial magnetic anisotropy parameter $D$, bias voltage $eV=\mu_{L}-\mu_{R}$, and Larmor frequency $\omega_L$ (magnetic field $B$), focusing on the influence of tuning the anisotropy parameter $D$ on charge transport properties of the system. 
In particular, it will be shown that the anisotropy parameter $D$ can contribute to controlling and reducing the noise power, especially via quantum interference effects, that occur at resonant conditions $\mu_{\xi}=\epsilon_i$, and manifest themselves as Fano-like peak-dip (dip-peak) features in the noise $S_{LL}$. 

In Fig.~3 the average charge current from the left lead $I_L$ and autocorrelation shot noise $S_{LL}$ are presented as functions of the magnetic anisotropy parameter $D$, for five different tilt angles $\theta$, while the Fano factor $F$ is shown in Fig.~4. 
The current and noise in Fig.~3 show the molecular quasienergy spectrum, with each dip, dip-peak, peak-dip, or step-like feature denoting the anisotropy parameter $D$ corresponding to the matching between level $\epsilon_i$ and chemical potential $\mu_{\xi}$.
The fano factor $F<1$, so the noise is sub-Poissonian. 
Both elastic tunneling, and inelastic processes involving absorption (emission) of an energy $\omega$ with longer electron dwell time on the molecular orbital, contribute to sub-Poissonian noise.
We see that the current, shot noise and consequently the Fano factor are constant for $\theta=0$ (orange, dotted lines) and $\theta=\pi/2$ (pink, dashed lines). If we look at the expression for the current, given by Eq.~(9), we notice that for $\theta=0$,  $\gamma=0$ as well, and the charge current dependence on $D$ vanishes. On the other hand, $S_{z}=0$ for $\theta=\pi/2$, hence the current does not depend on the anisotropy parameter $D$ either. For the tilt angle $\theta=\pi/3$ (green line in  Figs.~3 and 4) and $D=-0.01312\,\epsilon_0$ (grid line), corresponding to $\mu_{L}=\epsilon_{2}=2.5\,\epsilon_0$ and $\epsilon_{1}=0.69\,\epsilon_0$, there is a dip-peak feature in the current, and a peak-dip feature in the shot noise and Fano factor. The position of the quasienergy levels $\epsilon_i$ with respect to $\mu_{\xi}$ for this set of parameters corresponds to Fig.~2(d), where we see the available tunneling processes between levels with energies $\epsilon_1$ and $\epsilon_2$, and 2(i) where the tunneling processes between levels with energies $\epsilon_3=-0.5\,\epsilon_0$ and $\epsilon_4=1.31\,\epsilon_0$ are shown.
There are two available spin-flip processes for electrons tunneling into $\downarrow$ level $\epsilon_{1}$($\epsilon_3$) from the left lead [green curved arrows in Figs.~2(d) and 2(i)], and one elastic tunneling process [black horizontal arrow in Fig.~2(d)]. For an electron from the left lead entering  $\uparrow$ level with energy $\epsilon_2$, one spin-flip process is available in Fig.~2(d)(magenta curved line), while one electron can jump into the left lead from $\uparrow$ level $\epsilon_2$, and elastic tunnelling to $\uparrow$ level $\epsilon_2$ ($\epsilon_4$) is available [black horizontal arrow in Fig.~2(d)]. As a result $I^{\uparrow}_{L}<I^{\downarrow}_{L}>0$.
   
\begin{figure}
	\includegraphics[height=5.5cm,keepaspectratio=true]{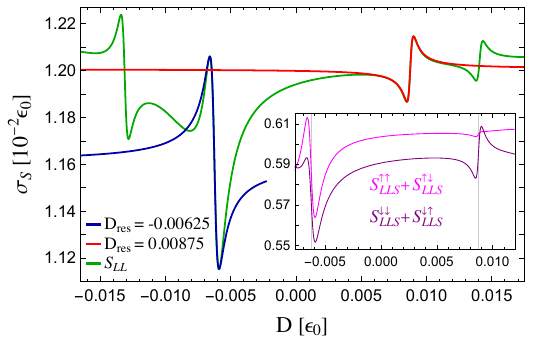}   
	\caption{Fano-like shape of the resonance profile $\sigma_S$ as a function of the uniaxial anisotropy parameter $D$, corresponding to noise $S_{LL}$ for $\theta=\pi/3$ in Fig.~\ref{fig: current_D}(b) (green line), for two resonant anisotropy parameters $D_{\rm res}=-0.00625\,\epsilon_0$ (blue line), and $D_{\rm res}=0.00875\,\epsilon_0$ (red line). Around $D_{\rm res}$ the shot noise $S_{LL}$ matches $\sigma_S$. The inset shows contributions of $S^{\uparrow\uparrow}_{LLS}+S^{\uparrow\downarrow}_{LLS}$ (pink line) and $S^{\downarrow\downarrow}_{LLS}+S^{\downarrow\uparrow}_{LLS}$ (purple line) to the resulting shapes of the resonance profiles in $S_{LL}$. The other parameters are the same as in Fig.~\ref{fig: current_D}.}\label{fig: Fano_resonance}
\end{figure}

\begin{figure*}
	\includegraphics[height=5.75cm,keepaspectratio=true]{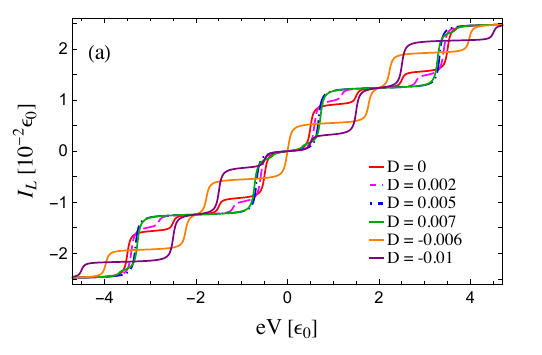}\,\,\,\includegraphics[height=5.75cm,keepaspectratio=true]{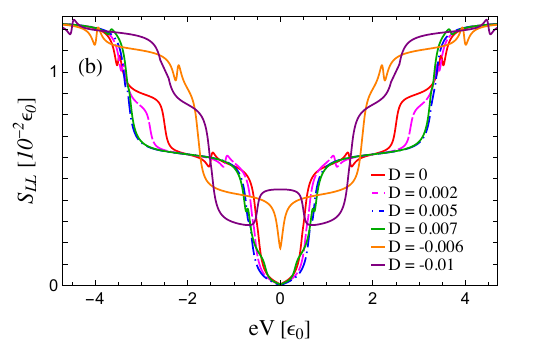}   
	\caption{(a) Charge current $I_L$ and (b) autocorrelation shot noise $S_{LL}$ as functions of the applied bias voltage $eV=\mu_{L}-\mu_{R}$ with $\mu_{L,R}=\pm eV/2$ and $\vec{B}=B\vec{e}_{z}$, for  different uniaxial magnetic anisotropy parameters $D$ at zero temperature. The other parameters are set to $\Gamma=\nobreak0.05\, \epsilon_{0},\,\Gamma_{L}=\Gamma_{R}=\Gamma/2,\,\omega_L=\nobreak0.5\,\epsilon_{0},\, J=\nobreak0.01\,\epsilon_{0},\, S=100$, and $\theta=\pi/3$.  All energies are given in the units of $\epsilon_{0}$. The peak-dip (dip-peak) features in the shot noise $S_{LL}$ as manifestations of the quantum interference effect, and steps in the charge current function $I_L$ and shot noise $S_{LL}$ correspond to resonances $\mu_{\xi}=\epsilon_{i}$, with $\xi=L,R$ and $i=1,2,3,4$.}\label{fig: current_eV}
\end{figure*}

\begin{figure}
	\includegraphics[height=5.7cm,keepaspectratio=true]{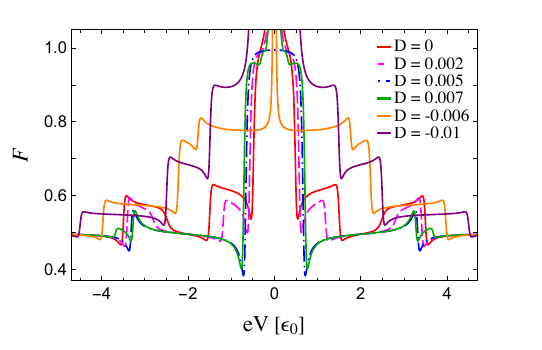}   
	\caption{Fano factor $F$  as a function of the applied bias voltage $eV=\mu_{L}-\mu_{R}$ with $\mu_{L,R}=\pm eV/2$ and $\vec{B}=B\vec{e}_{z}$, for different uniaxial magnetic anisotropy parameters $D$ at zero temperature. The other parameters are set to $\Gamma=\nobreak0.05\, \epsilon_{0},\,\Gamma_{L}=\Gamma_{R}=\Gamma/2,\,\omega_L=\nobreak0.5\,\epsilon_{0},\, J=\nobreak0.01\,\epsilon_{0},\, S=100$, and $\theta=\pi/3$.  All energies are given in the units of $\epsilon_{0}$.}\label{fig: Fano_eV}
\end{figure}

\begin{figure*}
	\includegraphics[height=5.7cm,keepaspectratio=true]{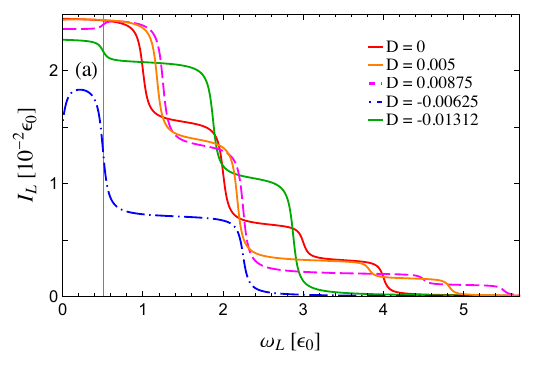}\,\,\,\,
	\includegraphics[height=5.7cm,keepaspectratio=true]{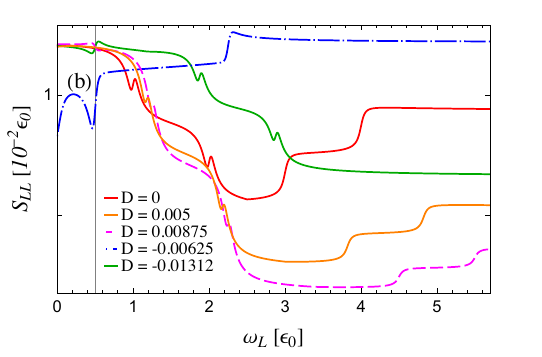}   
	\caption{(a) Charge current $I_L$ and (b) autocorrelation shot noise $S_{LL}$ as functions of the Larmor frequency $\omega_L$ for different uniaxial magnetic anisotropy parameters $D$. All plots are obtained at zero temperature with $\vec{B}=B\vec{e}_{z}$. The chemical potentials of the leads are equal to $\mu_{R}=0$ and $\mu_{L}=\nobreak2.5\,\epsilon_0$, except for $D=-0.00625\,\epsilon_0$, where $\mu_{L}=1.125\,\epsilon_0$. The other parameters are set to $\Gamma=\nobreak0.05\, \epsilon_{0},\,\Gamma_{L}=\Gamma_{R}=\Gamma/2,\, J=\nobreak0.01\,\epsilon_{0},\, S=100$, and $\theta=\pi/3$.  All energies are given in the units of $\epsilon_{0}$. All steps in the current $I_L$, and the corresponding steps and dip-peak features in the noise $S_{LL}$, denote a resonance $\mu_{\xi}=\epsilon_i$, with $\xi=L,R$. A distinct dip in the noise $S_{LL}$ and a significant drop of the corresponding current appear for $D=-0.00625\,\epsilon_0$ (blue dot-dashed line) around $\omega_{L}=0.5\,\epsilon_0$ (grid line), due to the double resonance: $\mu_{R}=\epsilon_3$ and $\mu_{L}=\epsilon_4$. }\label{fig: current_omega_L}
\end{figure*}
\begin{figure} 
	\includegraphics[height=5.7cm,keepaspectratio=true]{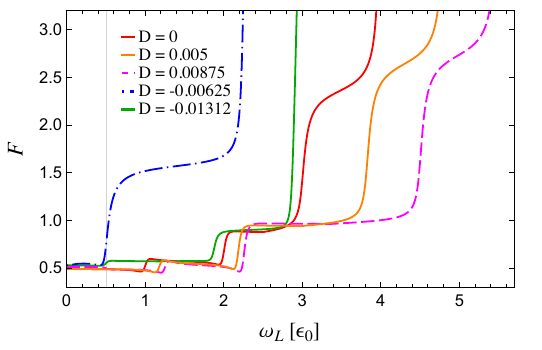}    
	\caption{Fano factor $F$ as a function of the Larmor frequency $\omega_L$ for different uniaxial magnetic anisotropy parameters $D$. All plots are obtained at zero temperature with $\vec{B}=B\vec{e}_{z}$. The chemical potentials of the leads are equal to $\mu_{R}=0$ and $\mu_{L}=\nobreak2.5\,\epsilon_0$, except for $D=-0.00625$ (blue dot-dashed line), where $\mu_{L}=1.125\,\epsilon_0$.
		The other parameters are set to $\Gamma=\nobreak0.05\, \epsilon_{0},\,\Gamma_{L}=\Gamma_{R}=\Gamma/2,\, J=\nobreak0.01\,\epsilon_{0},\, S=100$, and $\theta=\pi/3$.  All energies are given in the units of $\epsilon_{0}$.}\label{fig: Fano_omega_L}
\end{figure}

Similarly as in the Fano effect,\cite{interf2024,Fano2024} the peak-dip (dip-peak) features are manifestations of the quantum interference between the states connected with spin-flip events, accompanied by an energy change $\omega$. 
Namely, e.g., in Fig.~2(d) an electron from the left lead can tunnel through a $\uparrow$ state via an elastic tunneling process (black horizontal arrow), or via inelastic spin-flip process from $\downarrow$ to $\uparrow$ state, with absorption of an energy $\omega$, before it tunnels out of the orbital (green curved lines). The electron can alternatively tunnel  through a $\downarrow$ state elastically (black horizontal line), or via spin flip from $\uparrow$ to $\downarrow$ state, with emission of an energy $\omega$ (pink curved lines).
The superposition of two electron wave functions along the two tunneling pathways, one elastic and the other inelastic, ending in the same final $\uparrow$ or $\downarrow$ state leads to constructive (peaks) or destructive (dips) quantum interference, manifested in the shot noise.
The asymmetric line shape in $S_{LL}$ resembles the shape of the asymmetric Fano resonance profile, described by the Fano formula.\cite{interf2024,Fano2024} Around a resonant anisotropy parameter $D_{\rm res}$, for $\mu_{\xi}=\epsilon_i$, with destructive interference corresponding to a minimum (dip) at $D_{\rm min}$, and constructive interference corresponding to a maximum (peak) at $D_{\rm max}$, the shot noise $S_{LL}$ matches the Fano-like shape given by
\begin{equation}
	\sigma_{S}(D)=C+A\frac{(D-D_{\rm res}+q\Gamma_{\rm res}/2)^2}{(D-D_{\rm res})^2+(\Gamma_{\rm res}/2)^2},\label{eq: Fano_D_res}
\end{equation}	
where $q=\pm 1$ is the asymmetry parameter. For $\omega>0$, $q=-1$, whereas for $\omega<0$, $q=1$. The width of the resonance is $\Gamma_{\rm res}=|D_{\rm max}-D_{\rm min}|$, $C$ represents the shot noise at $D_{\rm min}$, $C=S_{LL}(D_{\rm min})$, while $A$ is half of the amplitude of the Fano-like shape, $A=[S_{LL}(D_{\rm max})-S_{LL}(D_{\rm min})]/2$. The asymmetry parameter $q$, represents the ratio of the probabilities of elastic and inelastic tunneling interfering pathways. Since $|q|=1$, both elastic and inelastic processes are equally probable, and $D_{\rm res}$ is located at equal distance from $D_{\rm min}$ and $D_{\rm max}$.

As the anisotropy parameter $D$ increases, the quasienergy level $\epsilon_3$ moves up the energy scale and enters the bias-voltage window around $D=-0.00625\,\epsilon_0$ for $\theta=\pi/3$ (another grid line in Figs.~3 and 4), with $\mu_{R}=\epsilon_{3}=0$, leading to the increase of the Fano factor since new quasienergy level becomes available for elastic or inelastic processes [see Figs.~2(g) and 2(h)]. Now, all levels satisfy $\mu_{R}<\epsilon_{i}<\mu_{L}$, leading to the enhancement of the current after its minimum value, with $I^{\uparrow}_{L}\approx I^{\downarrow}_{L}$>0, and the most prominent peak-dip feature in $S_{LL}$ and $F$, where the dips represent their minimum values.
The Fano-like shape of the resonance profile, $\sigma_{S}(D)$, given by Eq.~(22), corresponding to $S_{LL}$ with $\theta=\pi/3$ [green line in Fig.~3(b)] for $D_{\rm res}=-0.00625\,\epsilon_0$ and $D_{\rm res}=0.00875\,\epsilon_0$ is presented in Fig.~5.
The noise $S_{LL}$ resembles the Fano-like resonance profile $\sigma_S$, around resonant anisotropy parameters $D_{\rm res}$, $S_{LL}\approx\sigma_S$ (blue line around $D_{\rm res}=-0.00625\,\epsilon_0$ and red line around $D_{\rm res}=0.00875\,\epsilon_0$ in Fig.~5). For $D_{\rm res}=-0.00625\,\epsilon_0$, the parameters of the Fano-like profile $\sigma_S$ are the following: $q=-1$, $D_{\rm min}\approx -0.0059\,\epsilon_0$, $D_{\rm max}\approx -0.0066\,\epsilon_0$, $\Gamma_{\rm res}\approx 0.0007\,\epsilon_0$, $C\approx 1.1154\times 10^{-2}\,\epsilon_0$, and $A\approx 4.5312\times 10^{-4}\,\epsilon_0$. 
The inset shows individual contributions of two interference profiles to Fano-like shapes in $S_{LL}$. The interference of electron waves propagating, one elastically through $\uparrow$ level ($\epsilon_4$) and the other inelastically with absorption of an energy $\omega$, and a spin flip from $\downarrow$ level ($\epsilon_3$) to $\uparrow$ level ($\epsilon_4$) [see Fig.~2(h)], is presented in the contribution $S^{\uparrow\uparrow}_{LLS}+S^{\uparrow\downarrow}_{LLS}$ (pink line, inset in Fig.~5). The interference of
pathways involving $\downarrow$ state ($\epsilon_3$) and a spin flip from $\uparrow$ state ($\epsilon_4$) to $\downarrow$ state ($\epsilon_3$) accompanied by the emission of an energy $\omega$ is presented in $S^{\downarrow\downarrow}_{LLS}+S^{\downarrow\uparrow}_{LLS}$ (purple line, inset in Fig.~5). In both lines $q=-1$, showing equal probabilities of elastic and inelastic spin-flip processes involving absorption (pink line) or emission (purple line), with the ratio of amplitudes of their resonance profiles around $1.186$.  

With further increase of the parameter $D$, $I_{L}$ and $S_{LL}$ approach constant values, and $F\approx 0.49$,\cite{fanohalf} in the region between second and third grid lines, until $D=0.00875\,\epsilon_0$ [green line in Fig.~3(b)], since all levels $\epsilon_i$ lie within the bias-voltage window. The quasienergy level $\epsilon_4$ moves down the energy scale with the increase od $D$, and for $D=0.00875\,\epsilon_0$ (grid line in Figs.~3 and 4), $\mu_{R}=\epsilon_{4}=0$, leading to a decrease of $I_{L}$, with $I^{\uparrow}_{L}> I^{\downarrow}_{L}$>0 [tunneling processes in Figs.~2(g) and 2(h)] and a dip-peak in $S_{LL}$. Around $D_{\rm res}=0.00875\,\epsilon_0$, $S_{LL}$ resembles the Fano-like resonance shape given by Eq.~(22) (red line in Fig.~5). The asymmetry parameter is now $q=1$ as a consequence of the reversed direction of the effective magnetic field $\vec{B}_{\rm eff}=(\omega/g\mu_{B})\vec{e}_z$ $(\omega<0)$. The $\downarrow$ level with energy $\epsilon_3$ lies above $\uparrow$ level with energy $\epsilon_4$ for $\omega<0$, $\epsilon_{3}>\epsilon_4$. The contribution of the interference between elastic tunneling pathway through $\uparrow$ level  ($\epsilon_4$) and inelastic emission pathway with a spin flip from $\downarrow$ level ($\epsilon_3$) to $\uparrow$ level ($\epsilon_4$), against the direction of the effective magnetic field $\vec{B}_{\rm eff}$ ($B_{\rm eff}<0$, $B>0$) is negligible (pink line, inset in Fig.~5), while the interference between elastic process through $\downarrow$ level ($\epsilon_3$) and inelastic absorption process accompanied with spin flip from  $\uparrow$ level ($\epsilon_4$) to  $\downarrow$ level ($\epsilon_3$) almost entirely participates in the formation of the Fano-like resonance profile (purple line, inset in Fig.~5).

As the parameter $D$ increases, the level $\epsilon_4$ leaves the bias-voltage window. At $D=0.01406\,\epsilon_0$, for $\theta=\pi/3$ (grid line in Figs.~3 and 4), $\mu_{L}=\epsilon_1$, before the level $\epsilon_1$ leaves the bias-voltage window with further increase of $D$, resulting in the final decrease in $I_{L}$, with $I^{\uparrow}_{L}> I^{\downarrow}_{L}$>0, [tunneling processes in Figs.~2(d) and 2(i)] and a dip-peak in $S_{LL}$. All the plots in Fig.~3, except for $\theta=0$ and $\theta=\pi/2$, show that both $I_{L}$ and $S_{LL}$ have minimum values around the value of $D$ that corresponds to the entrance of all $\epsilon_i$ into the bias-voltage window. 
Both $I_{L}$ and $S_{LL}$ are saturated at high anisotropy $\lvert D\rvert\gg\omega_{L}/2S_z$, and $F\approx 0.49$. In this case only two levels, $\epsilon_1$ and $\epsilon_4$ (for $D<0$, $\omega>0$), or $\epsilon_2$ and $\epsilon_3$ (for $D>0$, $\omega<0$), lie between $\mu_{L}$ and $\mu_{R}$, and more energy is needed to flip an electron spin to the direction of the effective magnetic field, so that its spin flip is energetically unfavourable. Besides, an electron dwell time on the molecular orbital during a spin-flip process also increases. Therefore, at a high anisotropy, $\lvert D\rvert\gg\omega_{L}/2S_z$, the dominant tunneling processes are elastic, with spin-resolved charge currents $I^{\uparrow}_{L}\approx I^{\downarrow}_{L}$ and correlations of currents with the same spins, $S^{\uparrow\uparrow}_{LL}\approx S^{\downarrow\downarrow}_{LL}$, while the probability of electron spin flip decreases, so that the contribution of the correlations of currents with opposite spins to noise $S^{\uparrow\downarrow}_{LL}+S^{\downarrow\uparrow}_{LL}\rightarrow 0$.

The average charge current $I_L$ as a function of the applied bias voltage $eV$ at zero temperature is plotted for six different values of anisotropy parameter $D$ in Fig.~6(a), where the bias voltage is varied such that $\mu_{L,R}=\pm eV/2$. 
As the bias voltage increases, a new channel available for electron transport, with an energy $\epsilon_i$ enters the bias-voltage window, resulting in a step increase in the current. Since the positions of the quasienergy levels $\epsilon_i$ depend on molecular spin anisotropy, for different values of $D$, the staircase current function will show steps at different values of $eV$. The shot noise of charge current $S_{LL}$ and Fano factor $F$ as functions of $eV$ are shown in Figs.~6(b) and 7.
The peak-dip (dip-peak) features, which occur due to the quantum interference, and steplike features in $S_{LL}$ and $F$, correspond to resonances $\mu_{\xi}=\pm eV/2=\epsilon_{i}$.
 Hence, they change their positions with the change of the magnetic anisotropy parameter $D$. For the set of parameters: $\omega_{L}=0.5\epsilon_0$, $\theta=\pi/3$, and $D=0.005\epsilon_0$ in Fig.~6 (blue, dot-dashed line), one obtains $\omega=\omega_{L}-2DS_{z}=0$, with only two transport channels, with energies $\epsilon=0.34\epsilon_0$ and $\epsilon\prime=1.66\epsilon_0$, available for the elastic tunnelling, denoted by the steps at $\pm eV/2=\epsilon$ and $\pm eV/2=\epsilon\prime$. In this case, the Fano factor is Poissonian, $F=1$, for $|eV|<2\epsilon$ (blue dot-dashed line in Fig.~7), since the transmission probability is very low, depending on the level broadening $\Gamma$, and the currents remain uncorrelated until the first channel available for electron transport appears.
 For $\omega\neq 0$, the average current $I_{L}$ is equal to zero at $eV=0$, but the noise $S_{LL}$ is contributed by the inelastic processes in which an electron flips its spin and absorbs an energy $\omega$, leading to the divergence of the Fano factor (see Fig.~7). The noise becomes sub-Poissonian ($F<1$) as soon as one of the levels $\epsilon_i$ enters the bias-voltage window, since the transmission probability increases. After all the levels $\epsilon_i$ enter the bias-voltage window [tunneling processes in Fig.~2(g)], the Fano factor becomes constant $F=1/2$.\cite{fanohalf} 
  
The average charge current $I_L$ and noise $S_{LL}$ as functions of the Larmor frequency $\omega_L$ at zero temperature, for several different values of the magnetic anisotropy parameter $D$, are shown in Figs.~8(a) and 8(b).
The corresponding Fano factor $F$ is presented in  Fig.~9. Here, the bias voltage is varied as $eV=\mu_L-\mu_R$, with $\mu_R=0$, while $\mu_L=2.5\,\epsilon_0$, except for the blue dot-dashed lines where $\mu_L=1.125\,\epsilon_0$. All steps in the current $I_L$, as well as steps and dip-peak features in the noise $S_{LL}$, correspond to a resonance $\mu_{\xi}=\epsilon_i$. 
At $\omega_L=0$, for $D=-0.00625\,\epsilon_0$ and $\mu_L=1.125\,\epsilon_{0}$ (blue dot-dashed lines in Fig.~8), both $I_{L}$ and $S_{LL}$ increase, since $\mu_L=\epsilon_1$, while $\epsilon_2=1.75\,\epsilon_0$, $\epsilon_3=0.25\,\epsilon_0$ and $\epsilon_4=0.875\,\epsilon_0$.
The spin-resolved charge currents $I^{\uparrow}_{L}>0$ and $I^{\downarrow}_{L}>0$.
 In Fig.~8(b), around $\omega_{L}=0.5\epsilon_0$ (grid line), one observes a small peak-dip for $D=0.00875\,\epsilon_0$ (pink line), as $\epsilon_4=\mu_{R}$ and quantum interference occurs between channels with energies $\epsilon_3$ and $\epsilon_4$, with $\epsilon_3>\epsilon_4$ ($\omega<0$), while for the anisotropy parameter $D=-0.01312\,\epsilon_0$ [green line in Fig.~8(b)] there is a dip-peak feature, as $\mu_L=\epsilon_2$, showing the impact of the quantum interference effect between channels with energies $\epsilon_1$ and $\epsilon_2$ [tunneling processes for all levels in Figs.~2(d) and 2(i)]. In both cases, the Fano factor is sub-Poissonian, $F<1$. Again, one can see that $F\rightarrow 1/2$ if all the quasienergy levels lie within the bias voltage window,\cite{fanohalf} e.g., for $D=0.005\,\epsilon_0$ and $\omega_{L}\le 1.18\,\epsilon_0$, where $\omega_{L}=1.18\,\epsilon_0$ corresponds to $\epsilon_{3}=\mu_{R}$ (orange line in Fig~9). 
 For $D=-0.00625\,\epsilon_0$, around $\omega_{L}=0.5\,\epsilon_0$, the double resonance occurs: $\mu_{R}=\epsilon_3$ and $\mu_{L}=\epsilon_4$, with $\epsilon_{1}=0.875\,\epsilon_0$ and $\epsilon_{2}=2\,\epsilon_0$ [see tunneling processes for all levels in Figs.~2(a) and 2(e)], resulting in a dip with higher magnitude in the noise $S_{LL}$ as a sign of destructive quantum interference between channels with energies $\epsilon_3$ and $\epsilon_4$, connected with spin-flip events [blue dot-dashed line in Fig.~8(b)]. The charge current $I_L$ shows a steplike decrease with $I^{\uparrow}_{L}<0$ and $I^{\downarrow}_{L}>0$, whereas in the Fano factor $F$ a step increase occurs around $\omega_{L}=0.5\,\epsilon_0$, leading to super-Poissonian noise, with $F>1$  (blue dot-dashed line in Fig.~9). For $D=-0.0625\,\epsilon_0$, and $\omega_{L}>0.5\,\epsilon_0$, the spin-resolved currents $I^{\uparrow}_L<0$ and $I^{\downarrow}_L>0$.
 
In Fig.~8(b), note that around $\omega_L$ corresponding to $\mu_{\xi}=\epsilon_i$, $S_{LL}$ either increases, or if the quantum interference occurs, decreases around Fano-like line shapes (except around $\omega_{L}=0.5$ for $D=-0.00625$).
Accordingly, for each value of $D$, around $\omega_L$ such that one remaining quasienergy level $\epsilon_i$ within the bias-voltage window is in resonance with the chemical potential of one of the leads, one notices the final step decrease in the current $I_L$, and the final step increase or a dip-peak feature in $S_{LL}$ (see Fig.~8). 
 For instance, at $\omega_L=2.25\,\epsilon_0$ and $D=-0.00625\,\epsilon_0$ (blue dot-dashed line), the remaining level within the bias-voltage window, $\epsilon_1=\mu_R$ [see tunneling processes for all levels in Figs.~2(b) and 2(c)], whereas, for $D=-0.01312\,\epsilon_0$ and $\omega_L=2.87\epsilon_0$ (green line in Fig.~8), the energy of the only level within the bias-voltage window equals $\epsilon_4=\mu_L$ [tunneling processes in Figs.~2(c) and 2(f)]. 
With further increase of $\omega_L$, all four levels $\epsilon_i$ lie out of the bias-voltage window, two below $\mu_{R}$: $\epsilon_{1}<\mu_{R}$ and $\epsilon_{3}<\mu_{R}$, and two above $\mu_{L}$: $\epsilon_{2}>\mu_{L}$ and $\epsilon_{4}>\mu_{L}$. Hence, the average current $I_{L}$ drops to zero, with nonzero spin-resolved currents $I^{\uparrow}_{L}=-I^{\downarrow}_{L}<0$, while $S_{LL}$ becomes a constant, due to the spin-flip absorption processes, and $F>1$, indicating the super-Poissonian noise. 
\begin{figure} 
	\includegraphics[height=5.75cm,keepaspectratio=true]{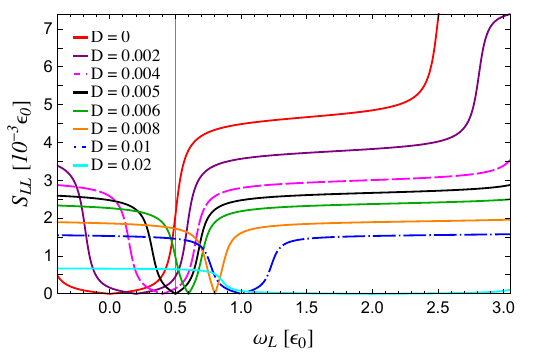}   
	\caption{Shot noise of charge current $S_{LL}$ as a function of the Larmor frequency $\omega_L$ for different uniaxial magnetic anisotropy parameters $D$ at zero-bias voltage. All plots are obtained at zero temperature with $\vec{B}=B\vec{e}_{z}$. The chemical potentials of the leads are equal $\mu_{L}=\mu_{R}=0.25\,\epsilon_0$.
			The other parameters are set to $\Gamma=\nobreak0.05\, \epsilon_{0},\,\Gamma_{L}=\Gamma_{R}=\Gamma/2,\, J=\nobreak0.01\,\epsilon_{0},\, S=100$, and $\theta=\pi/3$.  All energies are given in the units of $\epsilon_{0}$. The noise $S_{LL}$ is suppressed for $\omega=0$, i.e., $D=\omega_L/2S_z.$}\label{fig: negative}
\end{figure}

In Fig.~10 the dependence of the autocorrelation noise $S_{LL}$ on frequency $\omega_L$ is plotted for several values of the anisotropy parameter $D$ and equal chemical potentials of the leads, $\mu_L=\mu_R=\mu=0.25\epsilon_0$, at zero temperature. 
Here, only absorption processes between levels connected with spin-flip events occur.
For isotropic molecular spin with $D=0$ (red line in  Fig.~10), the shot noise $S_{LL}$ is an even function of $\omega_L$ at zero-bias conditions $eV=0$, $S_{LL}(\omega_L)=S_{LL}(-\omega_L)$,\cite{we2018} whereas if the molecular spin is anisotropic ($D\neq 0$), the shot noise is an even function of the frequency $\omega$, $S_{LL}(\omega)=S_{LL}(-\omega)$.
In all the plots each steplike increase or decrease corresponds to $\mu=\epsilon_i$ (not all are shown). For instance, if we take Larmor frequency $\omega_L=0.5\,\epsilon_0$ (marked by a vertical grid line), there is a step-like increase for the magnetic anisotropy parameter $D=0$ (red line) since the chemical potential $\mu$ is in resonance with quasienergy level $\epsilon_3$, $\mu=\epsilon_3$. Similar step for $D=0.002\,\epsilon_0$ occurs around $\omega_L=0.586\,\epsilon_0$, where $\mu=\epsilon_3$ (purple line in Fig.~10). For $\omega_L=0.5\,\epsilon_0$ and $D=0.005\,\epsilon_0$, the resulting $\omega=0$, with only two molecular levels, positioned above $\mu$, at $\epsilon=0.34\,\epsilon_0$ and $\epsilon^{\prime}=1.66\,\epsilon_0$, and neither elastic nor spin-flip processes occur. The shot noise $S_{LL}$ monotonically decreases around $\omega_L=0.5\,\epsilon_0$, drops to zero at $\omega_L=0.5\,\epsilon_0$ (intersection between grid line and black line), since spin-resolved correlations $S^{\sigma\sigma'}_{LL}=0$, and then monotonically increases. Similarly, the charge current noise $S_{LL}$ is equal to zero at $\omega_L=0.6\,\epsilon_0$, for $D=0.006\epsilon_0$ (green line), at $\omega_L=0.8\,\epsilon_0$, for $D=0.008\,\epsilon_0$ (orange line), and at $\omega_L=\epsilon_0$, for $D=0.01\,\epsilon_0$ (blue dot-dashed line).
For $D=0.006\,\epsilon_0$, $\mu=\epsilon_4$ at $\omega_L=0.5\epsilon_0$ and one observes a step-like decrease (intersection between grid line and green line in Fig.~10). With further increase of the magnetic anisotropy parameter $D$, the noise $S_{LL}$ decreases (cyan line in Fig.~10), and for a sufficiently large $D$, drops to zero, as $S^{\uparrow\downarrow}_{LL}+S^{\downarrow\uparrow}_{LL}\rightarrow 0$ and $S^{\uparrow\uparrow}_{LL}=S^{\downarrow\downarrow}_{LL}\rightarrow 0$.
\begin{figure} 
	\includegraphics[height=5.75cm,keepaspectratio=true]{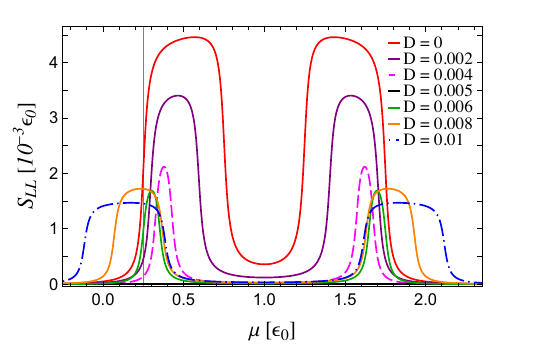}   
	\caption{Shot noise of charge current $S_{LL}$ at zero bias-voltage conditions, as a function of the chemical potential of the leads $\mu=\mu_{L}=\mu_{R}$, for different uniaxial magnetic anisotropy parameters $D$. All plots are obtained at zero temperature with $\vec{B}=B\vec{e}_{z}$.
			The other parameters are set to $\Gamma=\nobreak0.05\, \epsilon_{0},\,\Gamma_{L}=\Gamma_{R}=\Gamma/2,\, J=\nobreak0.01\,\epsilon_{0},\, S=100$, $\omega_{L}=0.5\,\epsilon_0$, and $\theta=\pi/3$.  All energies are given in the units of $\epsilon_{0}$. The shot noise $S_{LL}$ is positive between levels connected with spin-flip absorption events for $\omega\neq 0$, i.e., $D\neq\omega_L/2S_z$.}\label{fig: chempot}
\end{figure}

The autocorrelation shot noise $S_{LL}$ as a function of chemical potential of the leads at zero-bias voltage conditions $\mu=\mu_L=\mu_R$, when the average charge current is equal to zero, for several values of the magnetic anisotropy parameter $D$, at zero temperature, is shown in Fig.~11. The Larmor frequency $\omega_L=0.5\,\epsilon_0$ and the tilt angle $\theta=\pi/3$. 
Since the contribution of the correlations between currents with the same spin $S^{\sigma\sigma}_{LL}$ is dominant, the noise $S_{LL}$ takes positive values for $\omega\neq 0$, i.e., for $D\neq \omega_{L}/2S_z$ in the regions between quasienergy levels connected with spin-flip absorption processes, taking into account the level broadening $\Gamma$. The grid line in Fig.~11 at $\mu=0.25\,\epsilon_0$, corresponding to the grid line in  Fig.~10 at $\omega_L=0.5\,\epsilon_0$, intersects with all the plots showing that the autocorrelation noise $S_{LL}$ is reduced with the increase of the magnetic anisotropy parameter $D$. 
The noise can vanish at zero-bias voltage conditions for $\omega=0$, and that is satisfied for $D=\omega_{L}/2S_z=0.005\,\epsilon_0$ (black line in Fig.~11).
In that case the spin-resolved correlations $S^{\sigma\sigma'}_{LL}=0$.
With further increase of the magnetic anisotropy parameter $D$, the frequency $\omega<0$, i.e., the direction of the precession is altered, the anisotropy parameter $D>\omega_{L}/2S_z$, and the shot noise $S_{LL}$ takes positive low values between quasienergy levels connected with spin-flip events (green, orange, and blue dot-dashed lines in Fig.~11), compared to the one at the highest positive frequency $\omega=\omega_L$, for the isotropic molecular spin (red line in in Fig.~11). In order to reduce the shot noise $S_{LL}$ at zero-bias voltage conditions, one needs to increase the magnetic anisotropy parameter $D$ and either slow down the precession of the molecular spin, or reverse the direction of the spin precession with respect to the Larmor precession. With further increase of the magnetic anisotropy parameter, for a sufficiently large anisotropy parameter $D$, the shot noise $S_{LL}$ becomes entirely suppressed and drops to zero. 
\\

\section{Conclusions}

In this paper, the characteristics of charge transport through a single molecular orbital in the presence of a precessing anisotropic molecular spin in a magnetic field, connected to two noninteracting metallic leads, was theoretically studied. The Larmor frequency is modified by a term with the uniaxial magnetic anisotropy parameter of the molecular spin, and the resulting precession is externally kept undamped. The expressions for charge current and current noise were obtained using the Keldysh nonequilibrium Green's functions technique.

The results show rich transport characteristics at zero temperature. The quantum interference effects between the states connected with precession-assisted inelastic tunnelling processes, involving absorption(emission) of an energy $\omega$ and a spin flip, manifest themselves as peaks (constructive interference) and dips (destructive interference) in the shot noise, matching the Fano-like line shapes, controlled by the anisotropy parameter and Larmor frequency. Each resonance between a chemical potential and an anisotropy dependent quasienergy level is visible in a transport measurement through characteristics such as steps, peaks and dips, and can be varied by tuning the anisotropy. The correlations between the same-spin (opposite-spin) currents are positive (negative), and are particularly interesting at zero-bias conditions since the resulting shot noise is positive for chemical potentials between quasienergy levels, which are connected by the inelastic absorption processes accompanied with a spin flip. 
Here, with the increase of the anisotropy parameter, the precession frequency decreases or the precession direction becomes altered, and the noise is reduced. Additionally, the anisotropy parameter can be adjusted to suppress the precession frequency, so that the resulting shot noise vanishes. It was shown that the charge current and shot noise can be controlled by a proper adjustment of the anisotropy parameter of the molecular magnet and reach their saturation if the anisotropy is large at nonzero bias voltage.

Taking into account that the charge transport in the given setup with the anisotropic magnetic molecule can be manipulated by the uniaxial magnetic anisotropy parameter of the molecular spin, and other parameters, the results of this study may be useful in the field of single-molecule electronics and spintronics. It might be useful for magnetic storage applications to study the charge- and spin-transport properties using a setup with a molecular spin modelled as a quantum object in the future.

\begin{acknowledgments}
	The author acknowledges funding provided by the Institute of Physics Belgrade, through the grant No: 451-03-68/2022-14/200024 of the Ministry of Education, Science, and Technological Development of the Republic of Serbia.
\end{acknowledgments}

\begin{widetext}
	
	\appendix*
	
	\section*{Appendix: Derivation of Floquet quasienergies}
	\renewcommand{\theequation}{A\arabic{equation}}
	\setcounter{equation}{0}
	
Here, the derivation of the Floquet quasienergies given by Eqs.~(11) and (12) is presented. A periodic Hamiltonian $\hat{H}(t)=\hat{H}(t+\mathcal{T})$, with a period $\mathcal{T}=2\pi/\omega$ can be written as a Fourier series
\begin{equation}
	\hat{H}(t)=\sum_{n=-\infty}^{\infty}\hat{H}^{(n)}e^{in\omega t}.\label{eq: Appendixeq1}
\end{equation}
According to the Floquet theorem,\cite{Floquet1,Floquet2,Floquet3,Floquet4} the so called quasienergy states
\begin{equation}
	\lvert\psi_{\alpha}(t)\rangle=e^{-i\epsilon_{\alpha} t}\lvert\phi_{\alpha}(t)\rangle,\label{eq: Appendixeq3}
\end{equation}
with quasienergies $\epsilon_{\alpha}$ and time-periodic functions $\lvert\phi_{\alpha}(t)\rangle=\lvert\phi_{\alpha}(t+\mathcal{T})\rangle$, represent solutions of the Schr\"{o}dinger equation
\begin{equation}
	\hat{H}(t)\lvert\psi(t)\rangle=i\hbar\frac{\partial}{\partial t}\lvert\psi(t)\rangle.\label{eq: Appendixeq2}
\end{equation}
Substituting Eq.~(A2) into Eq.~(A3), one obtains\cite{Floquet1,Floquet2,Floquet3,Floquet4}
\begin{equation}
	\hat{\mathcal{H}}(t)\lvert\phi_{\alpha}(t)\rangle=\epsilon_{\alpha}\lvert\phi_{\alpha}(t)\rangle,
\end{equation}
where $\hat{\mathcal{H}}(t)=\hat{H}(t)-i\partial/\partial t$. Using an orthonormal basis $\{\lvert\alpha\rangle\}$ for the Hilbert space in which the Hamiltonian $\hat{H}(t)$ can be represented , the quasienergy state $\lvert\psi_{\alpha}(t)\rangle$ can be expanded as a Fourier series, while the quasienergies $\epsilon_\alpha$ can be calculated as the eigenvalues of the Floquet Hamiltonian $\hat{H}^{\rm F}$, 
\begin{equation}
	{\rm det}\lvert\hat{H}^{\rm F}-\epsilon\hat{I}\lvert=0.
\end{equation}
The matrix elements of $\hat{H}^{\rm F}$ in the orthonormal basis $\{\lvert\alpha n\rangle\}$, with $\lvert\alpha n\rangle=\lvert\alpha\rangle\otimes\lvert n\rangle$ representing products $\lvert\alpha\rangle e^{in\omega t}$, with $n\in\mathbb{Z}$, are given by\cite{Floquet2} 
\begin{equation}
	\langle\alpha n\lvert\hat{H}^{\rm F}\lvert\beta m\rangle=\langle\alpha\lvert\hat{H}^{(n-m)}\lvert\beta\rangle+n\omega\delta_{\alpha\beta}\delta_{nm},
\end{equation}
which can be further expressed as
\begin{equation}
	H^{\rm F}_{\alpha n,\beta m}=H^{(n-m)}_{\alpha\beta}+n\omega\delta_{\alpha\beta}\delta_{nm}.\label{eq: Appendixeq8}
\end{equation}
The infinite number of quasienergies, $\epsilon_\alpha+n\hbar\omega$ can be obtained for each quasienergy state $\lvert\psi_{\alpha}(t)\rangle$, so that one only needs to calculate quasienergies within one interval, e.g., $[0,\omega)$.

In the Hilbert space spanned by the eigenvectors of operator $\hat{s}_{z}$, $\lvert\uparrow\rangle=\lvert 1\rangle$ and $\lvert\downarrow\rangle=\lvert 2\rangle$, the time-periodic Hamiltonian of the molecular orbital $\hat{H}_{\rm MO}(t)=\hat{H}_{\rm MO}(t+2\pi/\omega)$ reads
	\begin{equation}
		\hat{H}_{\rm MO}(t)=\lambda_{1}\lvert 1\rangle\langle 1\rvert+\lambda_{2}\lvert 2\rangle\langle 2\rvert+\frac{JS_{\bot}}{2}e^{-i\omega t}\lvert 1\rangle\langle 2\rvert+\frac{JS_{\bot}}{2}e^{i\omega t}\lvert 2\rangle\langle 1\rvert,\label{eq: Appendixeq10}
	\end{equation}
with $\lambda_{1}=\epsilon_{0}+(\omega_{L}+JS_{z})/2$, $\lambda_{2}=\epsilon_{0}-(\omega_{L}+JS_{z})/2$ and $\omega=\omega_{L}-2DS_z$. The Hamiltonian of the molecular orbital expressed as a Fourier series 
\begin{equation}
	\hat{H}_{\rm MO}(t)=\sum_{n=-\infty}^{\infty}\hat{H}_{\rm MO}^{(n)}e^{in(\omega_{L}-2DS_{z}) t},\label{eq: Appendixeq11}
\end{equation}
has the following nonzero components, calculated using Eq.~(A7):
\begin{align}
	\hat{H}^{(0)}_{\rm MO}&=\lambda_{1}\lvert 1\rangle\langle 1\rvert+\lambda_{2}\lvert 2\rangle\langle 2\rvert,\label{eq: Appendixeq12}\\
	\hat{H}^{(-1)}_{\rm MO}&=\frac{JS_{\bot}}{2}e^{-i(\omega_{L}-2DS_{z}) t}\lvert 1\rangle\langle 2\rvert,\label{eq: Appendixeq13}\\
	\hat{H}^{(1)}_{\rm MO}&=\frac{JS_{\bot}}{2}e^{i(\omega_{L}-2DS_{z}) t}\lvert 2\rangle\langle 1\rvert.\label{eq: Appendixeq14}
\end{align}
According to Eq.~(A7) the matrix elements of the Floquet Hamiltonian can be calculated as
\begin{align}
	H^{\rm F}_{1n,1m}&=[\lambda_{1}+n(\omega_{L}-2DS_{z})]\delta_{nm},\\
	H^{\rm F}_{1n,2m}&=\frac{JS_\bot}{2}\delta_{n,m-1},\\
	H^{\rm F}_{2n,1m}&=\frac{JS_\bot}{2}\delta_{n,m+1},\\
	H^{\rm F}_{2n,2m}&=[\lambda_{2}+n(\omega_{L}-2DS_{z})]\delta_{nm}.
\end{align}
Since the Floquet Hamiltonian is block diagonal, it is enough to write one block,
\begin{equation}
	\left (
	\begin{array}{cc}
		\lambda_{1}+(n-1)(\omega_{L}-2DS_{z}) &  JS_{\bot}/2\\
		\vspace*{0.1cm}\\
		JS_{\bot}/2 & \lambda_{2}+n(\omega_{L}-2DS_{z})
	\end{array}
	\right ).
\end{equation}
The Floquet quasienergies $\epsilon_1$ and $\epsilon_3$ are the eigenvalues of the block for $n=0$, given by
\begin{equation}
	\epsilon_{1,3}=\epsilon_{0}-\frac{\omega_{L}}{2}+DS_{z}\pm\sqrt{D(D+J)S^{2}_{z}+\bigg (\frac{JS}{2}\bigg )^2},
\end{equation}
while the quasienergies $\epsilon_{2}=\epsilon_{1}+\omega$ and $\epsilon_4=\epsilon_{3}+\omega$ are the eigenvalues of the neighboring block with diagonal matrix elements shifted by $\omega$.
	
\end{widetext}

\end{document}